\magnification=\magstep1
\medskip
\centerline{\bf Causality and Electromagnetic Transmissions 
Through Materials}
\medskip
\centerline{V. Kidambi and A. Widom}
\centerline{Physics Department, Northeastern University, Boston MA 02115 }
\bigskip 
\centerline{\it ABSTRACT}
\medskip
There have been several experiments which hint at evidence for 
superluminal transport of electromagnetic energy through a material
slab. On the theoretical side, it has appeared evident that 
acausal signals are indeed possible in quantum electrodynamics.  
However, it is unlikely that superluminal signals can be 
understood on the basis of a purely classical electrodynamic signals 
passing through a material. The classical and quantum  
theories represent quite different views, and it is the quantum view 
which may lead to violations of Einstein causality.  
\medskip
\vskip 1.5cm
\centerline{\bf 1. Introduction}
\medskip
The difference between the {\it classical} notion of an electromagnetic  
wave, and the {\it quantum} notion of a photon can become easily blurred. 
The Maxwell wave of the classical electromagnetic theory, when Fourier 
transformed from space and time $({\bf r},t)$ into momentum space and time 
$({\bf k},t)$, mathematically produces the Schr\"odinger equation for 
the photon[1]. Thus, experiments which would have been completely 
understood by Maxwell from a classical electromagnetic viewpoint, are 
sometimes mistakenly held as evidence for a purely quantum mechanical 
effect. 

Our purpose is to discuss whether or not it is possible to propagate 
electromagnetic information at speeds faster than light speed, especially 
within the {\it classical} electrodynamic theory of continuous media[2]. 
Care shall be taken to consider the classical electrodynamic theory as 
different from the quantum electrodynamic theory. The profound difference 
between classical and quantum electrodynamics is already evident if 
one merely considers the vacuum propagators[3] of the two pictures 
(in the Feynman gauge); (i) If in the classical theory one employs 
the retarded vacuum photon propagator, 
$$
D_{retarded}({\bf r}-{\bf r}^\prime ,t-t^\prime )=
{
\delta \big(c(t-t^\prime )-|{\bf r}-{\bf r}^\prime |\big)
\over |{\bf r}-{\bf r}^\prime |
}\ \ \ \ (classical). \eqno(1)
$$ 
then the classical electromagnetic signal in the vacuum moves strictly 
on the light cone. (ii) If in the quantum theory one employs the Feynman 
propagator, 
$$
D_{Feynman}({\bf r}-{\bf r}^\prime ,t-t^\prime )=
{i\over \pi}\Big(
{
1\over |{\bf r}-{\bf r}^\prime |^2-c^2(t-t^\prime )^2+i0^+
}\Big)\ \ \ \ (quantum), \eqno(2)
$$
then the signal might move on the light cone (forward and backward in time),  
but the signal might also move off the light cone. The Feynman propagator 
describes a photon which can move any way it pleases. 

In what follows, we discuss classical solutions of the 
retarded type, similar to Eq.(1), but in the presence of 
material media. We discuss whether or not causality can 
be violated in the form of a superluminal signal within 
the context of locally retarded material response 
functions. We conclude that causality from the 
strictly classical electromagnetic viewpoint remains in tact.
Our mathematical methods of proof avoid all of the many different 
definitions of signal velocities and ``barrier'' transit times 
which fill much of the literature in this subject. For us, the system 
is causal if and only if the ``output'' depends only on what the 
``input'' has done in the past. If the ``output'' depends on what the 
``input'' will do in the future, then the system is acausal. 
\medskip 
\centerline{\bf 2. Material Polarization}
\medskip
Consider a material described by the displacement field 
$$
{\bf D}({\bf r},t)={\bf E}({\bf r},t)+4\pi {\bf P}({\bf r},t), \eqno(4)
$$
wherein the polarization response of the material is modeled as a linear  
retarded response to the electric field which is local in space but 
non-local in time[4], 
$$
{\bf P}({\bf r},t)=\int_{0}^\infty {\bf E}({\bf r},t-s)d{\cal F}(s). 
\eqno(5)
$$ 
Eq.(5) corresponds to the case of a frequency dependent dielectric 
response function 
$$
\epsilon (\zeta )=1+4\pi \int_0^\infty e^{i\zeta t}d{\cal F}(t), 
\ \ \ ({\cal I}m\ \zeta > 0),\eqno(6)
$$
which is analytic in the upper half frequency plane[5], 
and obeys the dispersion relation 
$$
\epsilon (\zeta )=1+{2\over \pi}\int_0^\infty 
{\omega {\cal I}m\ \epsilon(\omega +i0^+) d\omega
\over (\omega^2-\zeta^2) }\ ,
\ \ \ ({\cal I}m\ \zeta > 0). \eqno(7)
$$
\medskip 
\centerline{\bf 3. Electromagnetic Propagation Through a Slab}
\medskip

Now let us consider the propagation of an electromagnetic 
wave passing through a slab of such a material. For an appropriate 
component $V$ of the electromagnetic field, and for (say) normal incidence 
along the $z$- axis, one expects for a material slab or ``barrier'' that   
$$
V(z,t)\to V_{in}\big(t-(z/c)\big)+ V_{reflected}\big(t+(z/c)\big),
\ \ {\rm as}\ \ z\to -\infty , \eqno(9)
$$
and that 
$$
V(z,t)\to V_{out}\big(t-(z/c)\big), \ \ {\rm as}\ \ z\to \infty .
\eqno(10)
$$
Eqs.(9) and (10) correspond to an incoming wave $V_{in}$, a reflected 
wave $V_{reflected }$, and a transmitted wave (through the barrier) 
$V_{out}$. 

The transmission through the barrier is evidently causal in the 
sense of classical special relativity if the outgoing wave at time $t$ 
depends {\it only} on how the incoming wave would have behaved at  
previous times $t-s$ with $s\ge 0$; Explicitly,  
$$
V_{out} (t)=V_{in} (t)-
\int_0^\infty V_{in} (t-s)d{\cal G}(s), 
\ \ \ ({\rm causal\ response}).\eqno(11)
$$ 
The causal Eq.(11) will hold true if and only if the 
transmission amplitude 
$$
\tau (\zeta )=1-\int_0^\infty e^{i\zeta t}d{\cal G}(t), 
\eqno(12)
$$
is analytic in the upper half frequency plane, ${\cal I}m\ \zeta > 0$. 
The point is that Eqs.(9) and (10) read, respectively, in the frequency 
domain as 
$$
V_\omega (z)\to e^{i\omega z/c}+\rho_\omega e^{-i\omega z/c},
\ \ {\rm as}\ \ z\to -\infty , \eqno(13)
$$
and 
$$
V_\omega (z)\to \tau_\omega e^{i\omega z/c}, \ \ {\rm as}\ \ z\to \infty ,
\eqno(14)
$$
where $\rho_\omega $ is the reflection amplitude and 
$\tau_\omega $ is the transmission amplitude.

The fraction of the electromagnetic wave intensity at 
frequency $\omega $ which passes through the material slab is 
given by $P(\omega )=|\tau_\omega |^2$, which leads to the 
following: 
\par \noindent
{\bf Central Theorem:} {\it If the transmission amplitude is the boundary 
value $\tau_\omega =\tau (\omega +i0^+)$ of a function $\tau (\zeta )$ 
of complex frequency $\zeta $ analytic in the upper half plane, i.e. if}  
$$
P(\omega )=\lim_{\sigma \to 0^+}\ |\tau (\omega +i\sigma )|^2 , 
\eqno(15)
$$
{\it then the transmission through the slab is causal in the sense of 
Eq.(11)}. 

It is thus not required to perform a detailed Fourier transform 
from the frequency domain to the time domain in order to prove causality  
or acausality. All that is required is an explicit demonstration 
that the transmission amplitude $\tau (\omega +i0^+)$ is indeed 
the boundary value of function $\tau (\zeta )$ analytic 
for ${\cal I}m\ \zeta > 0$. The rest follows from the mathematical 
nature of the Fourier transform according to general theorems proved by 
Titchmarsh[6].

To see what is involved, consider a slab of material of thickness $L$ 
with a dielectric response function $\epsilon (\zeta )$ which obeys 
Eq.(7). If $\omega {\cal I}m\ \epsilon (\omega +i0^+)\ge 0$, then the 
complex index of refraction $\eta (\zeta )$ obeys a similar 
dispersion relation with $\omega{\cal I}m\ \eta (\omega +i0^+)\ge 0$, 
$$
\eta (\zeta )=\sqrt{\epsilon (\zeta )}=
1+{2\over \pi}\int_0^\infty 
{\omega {\cal I}m\ \eta(\omega +i0^+) d\omega
\over (\omega^2-\zeta^2) }\ ,
\ \ \ ({\cal I}m\ \zeta > 0). \eqno(16)
$$
The transmission amplitude obeys 
$$
\tau (\zeta )=
{
2\eta (\zeta ) e^{i\zeta L/c}
\over 
\big(1+\eta (\zeta )^2\big)\cos(\zeta \eta (\zeta )L/c)
+2i\eta (\zeta )\sin(\zeta \eta(\zeta)L/c)
}\ .
\eqno(17)
$$
Since Eqs.(16) and (17) imply 
that $\tau (\zeta )$ is analytic in the upper half complex frequency 
plane, then Eq.(11) holds true and the transmission through the slab 
is strictly causal in the classical electrodynamic theory. The out going 
wave depends on what the incoming wave would have been doing in the 
past if the slab were not present. Note that the inequality 
$\omega {\cal I}m\ \epsilon (\omega +i0^+)\ge 0$ is essential 
for the classical proof of causality.
\medskip
\centerline{\bf 4. Conclusions}
\medskip

There is a very large literature on superluminal group velocities[7] 
and delay times[8] for electromagnetic transmission through slabs of 
materials. Much of this work uses many different definitions of  
velocities and delay times. Some work hints at the notion that there is 
a superluminal transmission of an electromagnetic signal. More 
often, the discussion exhibits a disclaimer which asserts that Einstein 
causality is not violated. For example, in one mathematical model[9] 
of microscopic objects with inverted populations, the complex index of 
refraction is given by    
$$
\eta_{Chiao} (\zeta )=
\sqrt{1-\Big({|f|\omega_p^2 \over \Omega^2-2i\zeta \gamma -\zeta^2}\Big)}. 
\eqno(18) 
$$     
If $\Omega^2 <|f|\omega_p^2$, then there is a ``cut'' for both $\eta $ 
and $\tau $ in the upper half frequency plane. For such a case, 
in accordance Eq.(17) and (18), as a mathematical model one finds 
a non-analytic $\tau (\zeta )$ for ${\cal I}m \zeta >0$ and 
this model {\it does indeed violate Einstein 
Causality!} No disclaimer needs to be be made for this 
result. However, the sum rule 
$$
{2\over \pi}\int_0^\infty 
\omega {\cal I}m\ \epsilon(\omega +i0^+) d\omega
=\omega_p^2>0, \eqno(19)
$$ 
holds true even if $\omega {\cal I}m\ \epsilon(\omega +i0^+)<0$  
for a limited bandwidth. Since the model in Eq.(18) obeys the 
inequality $\omega {\cal I}m\ \epsilon(\omega +i0^+)<0$ for all 
$\omega $, the model violates the sum rule Eq.(19) and is not entirely 
physical.

There is presently the possibility of building mathematical models with 
negative noise temperatures in finite bandwidth which genuinely violate 
Einstein causality. Some examples model ``inverted populations'' and the 
microscopic mechanism for the causality violation is to be found in 
quantum mechanics. An atom or molecule in an excited state can scatter 
a photon in the following manner: Firstly (in time), the excited atom may  
emit the outgoing photon, and secondly (in time) the ``virtual'' 
ground state atom may absorb the incoming photon. The total scattering 
is elastic and energy conserving. This is only true over times long 
with respect to the the duration of the scattering event.

Typical of definitions often employed for time delay in the literature 
(when an electromagnetic signal passes through a material slab) is 
that 
$$
t_{delay}(\omega )=\Big({d\theta (\omega )\over d\omega }\Big),  
\eqno(20)
$$
where 
$$
\tau_\omega =\sqrt{P(\omega )}e^{i\theta (\omega )}. \eqno(21)
$$
The criteria of a negative delay time $t_{delay}(\omega )<0$ or of 
group velocity $v=(d\omega /dk)>c$ are not reliable for 
studying acausality in experiments[10-12]. Such criteria often amount 
to little more than approximate Fourier transformations. Our central point 
is that criteria such as Eq.(11), take place strictly in the time domain. 
On this more strict basis, acausality is theoretically possible in 
mathematical models but has not yet been definitively observed.

\vskip .5cm
\centerline{\bf REFERENCES}
\bigskip

\item{[1]} A. I. Akhiezer \& V. B. Berestetskii, {\it Quantum Electrodynamics}, 
Interscience, (1965).

\item{[2]} L. D. Landau and E. M. Lifshitz {\it Electrodynamics of Continous 
Media}, Pergamon Press, Oxford (1975).

\item{[3]} R. P. Feynman, {\it Phys. Rev.} {\bf 76}, 769 (1949).

\item{[4]} L. D. Landau and E. M. Lifshitz, {\it op. cit.}, p. 249.

\item{[5]} L. D. Landau and E. M. Lifshitz, {\it op. cit.}, pp. 256-262.

\item{[6]} E. C. Titchmarsh, {\it Introduction to the Theory of Fourier 
Integrals}, 2nd. ed., p. 44, Oxford Clarendon Press, Oxford (1967). 

\item{[7]} R. Y. Chiao, E. Bolda, J. Boyce, J. C. Garrison and M. W. Mitchell,
{\it Quantum and Semiclassical Optics} {\bf 7} , 279 (1995).

\item{[8]} V. S. Olkhovsky and E. Recami, {\it Phys. Report} {\bf 214}, 339 
(1992). 

\item{[9]} R. Y. Chiao and J. Boyce, {\it Phys. Rev. Lett.} {\bf 73}, 3383, 
(1994).

\item{[10]} N.~G.~Basov, R.~V.~Ambartsumyan, V.~S.~Zuev, P.~G.~Kryukov and 
V.~S. Letokhov, {\it Soviet Physics JETP}, {\bf23}, 1, p.16 (1966).

\item{[11]} A.~M.~Steinberg, P.~G.~Kwiat and R.~Y.~Chiao {\it Phys. Rev. Lett.} 
{\bf 71}, 708 (1993).

\item{[12]} Ch.~Spielmann, R.~Szip\"ocs, A.~Sting and F.~Krausz, 
{\it Phys. Rev. Lett} {\bf 73}, 2308 (1994).  

\bye